\documentclass[conference]{IEEEtran}

\usepackage[cmex10]{amsmath}

\usepackage{cite}
\usepackage{graphicx}
\usepackage[T1]{fontenc}
\usepackage[utf8]{inputenc}
\usepackage{color, colortbl}
\usepackage{cite}
\usepackage{mathtools}
\usepackage{amsmath,bm}
\usepackage{amsfonts,bbold}
\usepackage{caption}
\usepackage[T1]{fontenc}
\usepackage[utf8]{inputenc}
\usepackage{subfigure}
\usepackage{hyperref}
\usepackage{tikz}
\usepackage{amsthm}
\usepackage{pgfplots}
\usepackage{wrapfig}
\usepackage[export]{adjustbox}
\usepackage{siunitx}
\usepackage[american,cuteinductors,smartlabels]{circuitikz}
\usepackage{float}
\usepackage{color}
\usepackage{graphicx}
\newtheorem{remark}{Remark}

\definecolor{light-gray}{gray}{0.96}

\begin{document}
    

\title{Beyond low-inertia systems: Massive integration of grid-forming power converters in transmission grids}

\author{
\IEEEauthorblockN{A. Crivellaro\textsuperscript{1,2}, A. Tayyebi\textsuperscript{1,3}, C. Gavriluta\textsuperscript{1}, D. Gro\ss\textsuperscript{3}, A. Anta\textsuperscript{1}, F. Kupzog\textsuperscript{1}, and F. D\"{o}rfler\textsuperscript{3}}
\IEEEauthorblockA{\textsuperscript{1}AIT Austrian Institute of Technology, Austria
\;\; \textsuperscript{2}University of Padova, Italy \;\; 
\textsuperscript{3}ETH Z\"{u}rich, Switzerland\\
Corresponding Author's E-mail: ali.tayyebi-khameneh@ait.ac.at
}
}

\maketitle

\begin{abstract}
As renewable sources increasingly replace existing conventional generation, the dynamics of the grid drastically changes, posing new challenges for transmission system operations, but also arising new opportunities as converter-based generation is highly controllable in faster timescales. This paper investigates grid stability under the massive integration of grid-forming converters. We utilize detailed converter and synchronous machine models and describe frequency behavior under different penetration levels. First, we show that the transition from $0\%$ to $100\%$ can be achieved from a frequency stability point of view. This is achieved by retuning power system stabilizers at high penetration values. Second, we explore the evolution of the nadir and RoCoF for each generator as a function of the amount of inverter-based generation in the grid. This work sheds some light on two major challenges in low and no-inertia systems: defining novel performance metrics that better characterize grid behaviour, and adapting present paradigms in PSS design.
\end{abstract}



\section{Introduction}


The demand for the reduction of the carbon footprint has led to an increasing integration of renewable sources. The replacement of conventional power plants, interfacing the grid via synchronous machines (SMs), with wind and solar generation results in significant changes in power system dynamics. Specifically, as these new converter-based sources replace SMs, the amount of rotational inertia in power systems decreases, accompanied with the loss of stabilizing control mechanisms that are present in SMs.

As a result of this transition, low-inertia power systems encounter critical stability challenges \cite{milano_foundations_2018}; EirGrid \& SONI, for instance, limited the instantaneous penetration of variable renewable energy sources to $55\%$ \cite{kroposki_achieving_2017} and recently increased the limit to $67\%$ and set a goal of $75\%$ of fuel-free generation \cite{Annual-Renewable-Constraint-and-Curtailment-Report-2017-V1}. As of now, certain grids need to preserve a minimum amount of inertia, which implies higher cost and hinders the penetration of renewable generation. New converter control strategies can potentially address these low-inertia system stability issues. These approaches can be split in two categories \cite{milano_foundations_2018,poolla_placement_2018}: grid-following control, where the converter \textit{follows} the measured frequency and voltage magnitude in the grid (via a synchronizing mechanism such as phase locked loop), and grid-forming control, where the converter \textit{defines} the voltage magnitude and frequency. Given the fact that the first strategy relies on the existence of a well-defined voltage waveform, it cannot fully replace the functionality of the SMs. In this work, we focus on grid-forming converters (GFCs) and their critical role in the transition towards a $100\%$ converter-based grid.

Different GFC control strategies have been proposed, such as 
droop control \cite{chandorkar_control_1993}, virtual synchronous machine \cite{zhong_synchronverters:_2011}, dispatchable virtual oscillator control \cite{colombino_global_2017} and matching control \cite{arghir_grid-forming_2017}, among others. To the best of our knowledge, various aspects of the integration of GFCs (e.g., the consequent gradual inertia reduction) in a realistic transmission grid model and in an electromagnetic transient (EMT)  simulation environment have not been thoroughly explored. 
In \cite{AT-DG-AA-FK-FD:19} and \cite{Uros2019}, different grid-forming and grid-following techniques have been tested in simple network models. However, these studies rely on the IEEE 9-bus system that lacks sufficient granularity and complexity to fully analyze the transition scenario to GFCs. 


The objective of this work is to explore the limits of GFC integration at the transmission level, using an EMT simulation of a realistic grid model that fully reflects the existing dynamics. Previous studies suggest that systems exhibit instability \cite{Uros2019,Annual-Renewable-Constraint-and-Curtailment-Report-2017-V1,lin2017stability} when the penetration of non-synchronous generation increases to roughly $70\%$. The study in \cite{lin2017stability} only considers grid-following converters, and the work in~\cite{Uros2019} shows that instability can be caused by adverse interactions of GFCs with the power system stabilizer (PSS) and automatic voltage regulator (AVR). Our work suggests that, given the right control strategies for converter-based generation, a minimum amount of inertia might not be required for grid operation, from a frequency stability perspective. Nonetheless, some controllers can no longer be agnostic to the amount of converter-based generation, as the grid dynamics varies drastically depending on the generation mix. Moreover, we question the suitability of standard frequency metrics, such as nadir and rate-of-change-of-frequency (RoCoF), for converter-dominated grids. This paper does not analyze other critical aspects in low-inertia systems such as voltage control, responsiveness to faults, etc.

The main contributions of this paper are: first, to show that, from a frequency stability perspective and for a particular grid, it is possible to transition from $0$ to $100\%$ converter-based generation. Second, we remark the need for PSS retuning based on the continuously changing amount of non-synchronous penetration. Third, we explore how the nadir and RoCoF, measured over different time windows, evolve as a function of the penetration of converter-based generation. These results expose new challenges that have been so far overlooked in the mainstream literature and calls for further research to address many open points, such as: are nadir and RoCoF still good descriptors of grid stability? how relevant are fast transients in frequency? can decentralized PSS structures provide adequate damping under different converter-dominated scenarios?

\section{Model description}
\label{model}
We start with the description of the grid, SMs, and GFCs models.

\subsection{Transmission grid model}
\label{modelGrid}
\begin{figure}[t!]
    \centering
    {\includegraphics[trim=4mm 1mm 0 1mm,clip,width=0.35\textwidth]{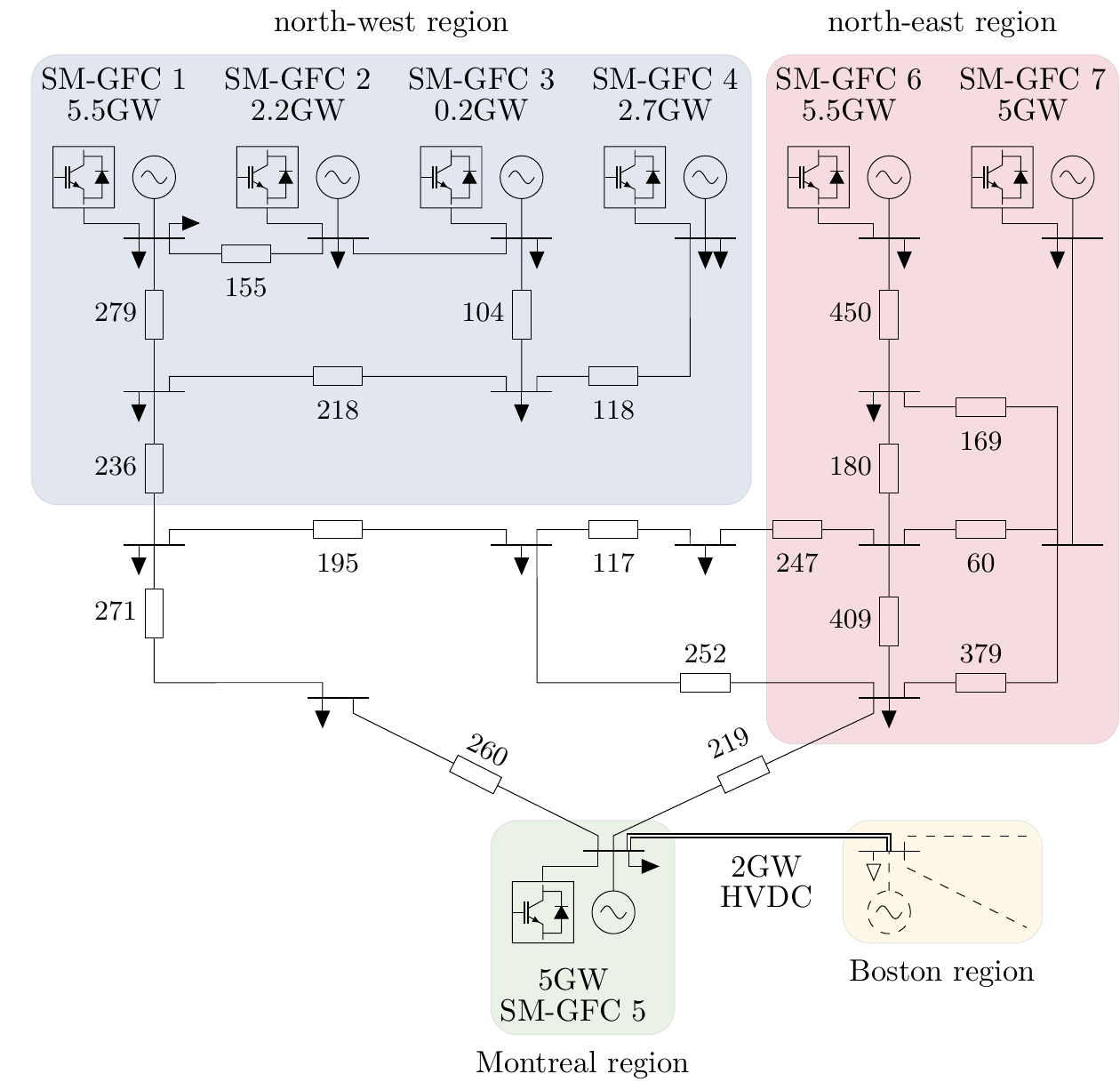}}
    \caption{Quebec grid model consisting of 7 generation nodes (line length in $\mathrm{km}$).}
    \label{fig:Quebec}
\end{figure}
In this work we adopt the transmission grid model from \cite{Quebec}, representing a simplified model of the Quebec region, with a total generation capacity of $26.2$ GVA consisting of seven SMs. This grid is characterized by three distinct regions which are interconnected via long transmission lines. 
Most of the generation can be found in the North and most of the load in the southern part. Even though the generation is mainly characterized by hydro-power plants, it has been selected for this study as the relevant information and the EMT simulation model are publicly available, as provided by Hydro-Quebec. Moreover, it has the right degree of complexity, i.e., being complex enough to explore the interactions of GFCs and SMs at different levels of inertia, and simple enough to understand the system behaviour. Specifically, inside the model there are $7$ SMs of different size, ranging from $5.5$ GW to $200$ MW. Each SM is represented by a set of $8$th order, $3$-phase dynamical model coupled with hydraulic turbine, governor, an AVR, and multi-band PSS (type 4B). Note that only primary frequency control is implemented in the SM model, and the droop constant of each SM is set to $5\%$. As explained later in the Section \ref{modelTransition}, we extend the model with an HVDC link of $2$ GW (existing in the original Hydro-Quebec grid but not in \cite{Quebec}), modeling a contingency that is independent of the penetration level of GFCs. For simplicity, we only consider constant impedance loads in our model. A simplified version of the grid model is depicted in Figure \ref{fig:Quebec}. 

\subsection{Grid-forming converter model}\label{modelConverter}
\begin{figure}[t!]
    \centering
    \begin{subfigure}
    {\includegraphics[width=0.35\textwidth]{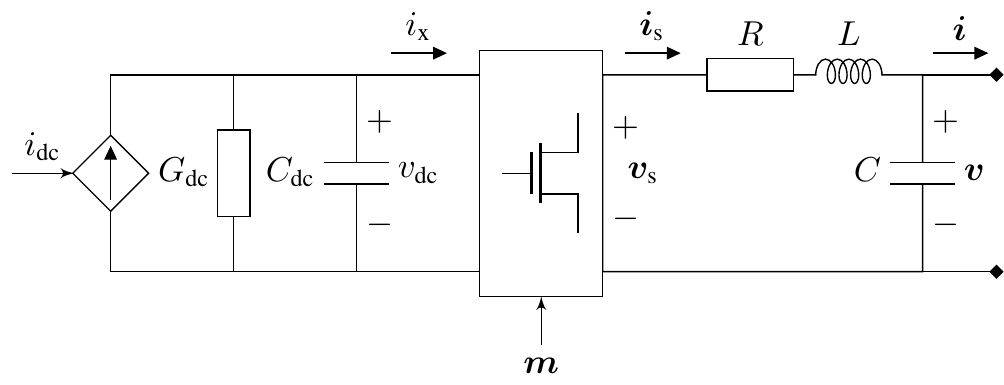}}
    \caption{Converter model (see \cite[Sec. II-A]{AT-DG-AA-FK-FD:19} for the parameters definition and further details of the converter modeling).\label{fig:converterModel}}    
    \end{subfigure}\vspace{5mm}
     \begin{subfigure}
    {\includegraphics[trim=6.5mm 0 0 0,clip,width=0.4\textwidth]{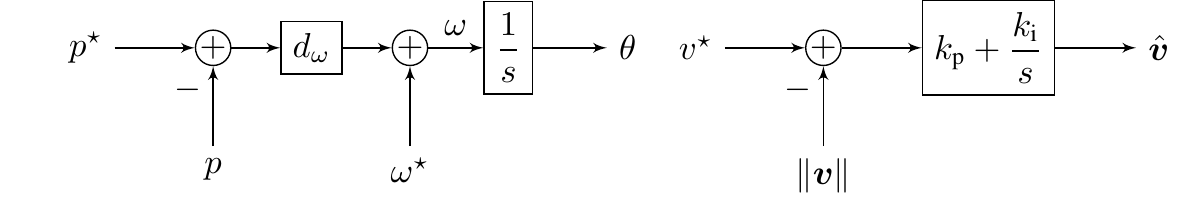}}
     \caption{Droop control block diagram [Sec. III-C]\cite{AT-DG-AA-FK-FD:19}.\label{fig:droop}} 
    \end{subfigure}
\end{figure}

The converter-based generation is implemented by means of two-level voltage source converters, stacked in parallel to form large-scale generation units \cite[Rem. 1]{AT-DG-AA-FK-FD:19}. The  converter DC energy source is a controllable current source, connected in parallel with a resistance (which models the DC losses) and the DC-link capacitance. The switching stage is modelled using a full-bridge 3-phase average model, AC output filter (see Figure~\ref{fig:converterModel}), and coupled to the medium voltage via a LV/MV transformer. Each converter is controlled as a grid-forming unit defining the angle, frequency and voltage. For simplicity, in this work we focus on grid-forming droop control (see \cite{chandorkar_control_1993}, \cite[Sec. III-C]{AT-DG-AA-FK-FD:19}). It is noteworthy that - under a realistic tuning and for a wide range of contingencies - other techniques such as virtual synchronous machine (VSM) \cite{zhong_synchronverters:_2011}, matching controlled GFCs \cite{arghir_grid-forming_2017} and dispatchable virtual oscillator control (dVOC) \cite{colombino_global_2017} exhibit similar behavior to that of the converters controlled by droop control\cite[Sec. IV]{AT-DG-AA-FK-FD:19}. The control block diagrams of the droop strategy appear in Figure \ref{fig:droop}. For the sake of compactness we refer the reader to \cite[Sec. III]{AT-DG-AA-FK-FD:19} and \cite{model} for further details on the converter control design. The droop gain is selected in order to provide the same load sharing capabilities as of the SMs in the system (see \cite[App. A]{AT-DG-AA-FK-FD:19} for details). Lastly, a proportional controller regulates the DC-link voltage via DC source current actuation \cite[Sec. III-B]{AT-DG-AA-FK-FD:19}. 
\subsection{Modeling the SM-GFC generation transition }
\label{modelTransition}
Given that the original model is an aggregated model, the generation transition in this study is carried out in a uniform, gradual way. Each SM is replaced by a collocated combination of SM and GFC, where the ratings of each generation unit is defined according to the penetration level\footnote{For a more granular grid model where each individual generator is included, it might be more realistic to represent this transition in a more discrete manner, where each SM is fully replaced by converter-based generation, one at a time.}. Formally speaking, the ratio of converter-based generation $\eta\in[0,1]$ is defined as:
\begin{equation}\label{eq:eta}
\eta = \frac{\sum_{i=1}^7 S_{\text{GFC}_i}}{\sum_{i=1}^7\left(S_{\text{GFC}_i}+S_{\text{SM}_i}\right)}. 
\end{equation}
where $S_{\text{SM}_i}$ denotes the rating of the \textit{i}-th SM in the combined model and $S_{\text{gfc}_i}$ denotes the rating for the \textit{i}-th converter. The individual ratings of the combined SM-GFC model replacing the original SM are then adjusted as a function of $\eta$:
\begin{subequations}\label{metricsEq}
\begin{align}
S_{\text{GFC}_i} &= \eta S_{\text{SM}_i}^0,\\
S_{\text{SM}_i} &= (1-\eta) S_{\text{SM}_i}^0,
\end{align}
\end{subequations}
with $S_{SM}^0$ being the rating of a given SM in the original model (with no GFCs, \textit{i.e.,} $\eta=0$). For $\eta=0$ (resp. $\eta=1$), the GFCs (resp. the SMs) are disconnected from the model.  The inertia time constant $H$ and the turbine time constant $\tau$ are kept constant regardless of the rating, as for a hydro-power plant it is more a function of the type of governor and turbine rather than the size\cite[Sec. 9.1]{kundur1994power}. Moreover, the original model delivered by Hydro-Quebec specifies identical parameter values for all the plants regardless of the size.  
\section{Results}
\label{results}
We start by defining the contingencies that will be considered. It is expected that, as SMs are being replaced by GFCs, the size of the worst contingency (typically the rated power of the largest SM in the grid) will become smaller, as generation becomes less coarse and more distributed. In our particular case study, this implies the worst contingency is the loss of the largest SM for low penetration levels (SM 1 or 6), and the HVDC link trip for high penetration levels. Nonetheless, for a fair comparison across different integration levels, we consider always the same contingency value for all values of $\eta$. Therefore, the worst contingency is chosen to be the simultaneous loss of the combined generation unit SM-GFC 1 (namely, the loss of $5.5$ {GW}  generation). For completeness, the disconnection of the HVDC link in the model will be considered as well in Sections \ref{resultsnadirRocof} and \ref{resultsAllGFC}. 

\subsection{PSS retuning for high penetration levels}
\label{resultsPSS}
It has been conjectured that the generation transition from a SM-dominated grid to a GFC-dominated one is challenging~\cite{Uros2019,DDPXM17,MBPQRUAMHVOQ19}. Indeed, we have observed in our initial results that, starting at $80\%$ GFC penetration, stability is lost. However, we found that re-tuning of the PSSs renders the system stable, at least from a frequency perspective. For $\eta\leq 0.7$, the system is stable under the original PSS structure (multiband PSS4B) and parameters, where all PSS blocks have the same parameters for all units. Roughly speaking, this type of PSS structure defines 3 different frequency bands and their corresponding lead-lag compensators. For the original PSS, these 3 frequency bands are set around 0.2Hz, 0.9Hz, and 12Hz, aimed at global, inter-area and local modes, respectively. For $0.8\leq\eta\leq 0.9$, the PSSs have been modified as follows: the second frequency range has been shifted to 1.2Hz, and the high frequency branch has been completely removed, to avoid having the corresponding lead-lag compensator acting on the existing GFC fast dynamics. Likewise, the gains of each branch have been reduced by a factor of 5. Based on this successful retuning, it can then be conjectured that, under the massive presence of GFCs, two aspects need to be considered: the PSS action might need to be reduced accordingly (but not fully removed); the PSS effect on high frequencies, where the response of GFCs is significant, can destabilize the system. Further analysis is needed to derive a more formal conclusion. We emphasize that there should be other re-tuning strategies that successfully stabilize the system, including the more natural choice of different PSS parameters for each SM.

Previous works~\cite{Uros2019,lin2017stability} already pointed at AVR and PSS regulators as the possible cause for instability at high values of $\eta$. 
Note  that the modified PSS tuning does not stabilize the system for $\eta\leq 0.7$. Finding a unique set of PSS parameters stabilizing the system for all penetration levels is a challenging task. Indeed, it is unclear whether such settings exist, as the system dynamics and oscillation modes drastically vary depending on the amount of GFCs present in the grid. In practice, it is undesirable to continuously retune the existing PSS controllers in a grid depending on the penetration level. Moreover, the real-time ratio of converter-based generation  (and its location) is not accurately known at the plant level, unless the transmission system operators discloses this information. Therefore, either novel robust, adaptive or more centralized PSS  structures would be required to guarantee stability independently of the amount of converter-based generation present in the grid.
\begin{figure}[t]
    \centering
    {\includegraphics[trim=2mm 2mm 2mm 0mm,clip,width=.45\textwidth]{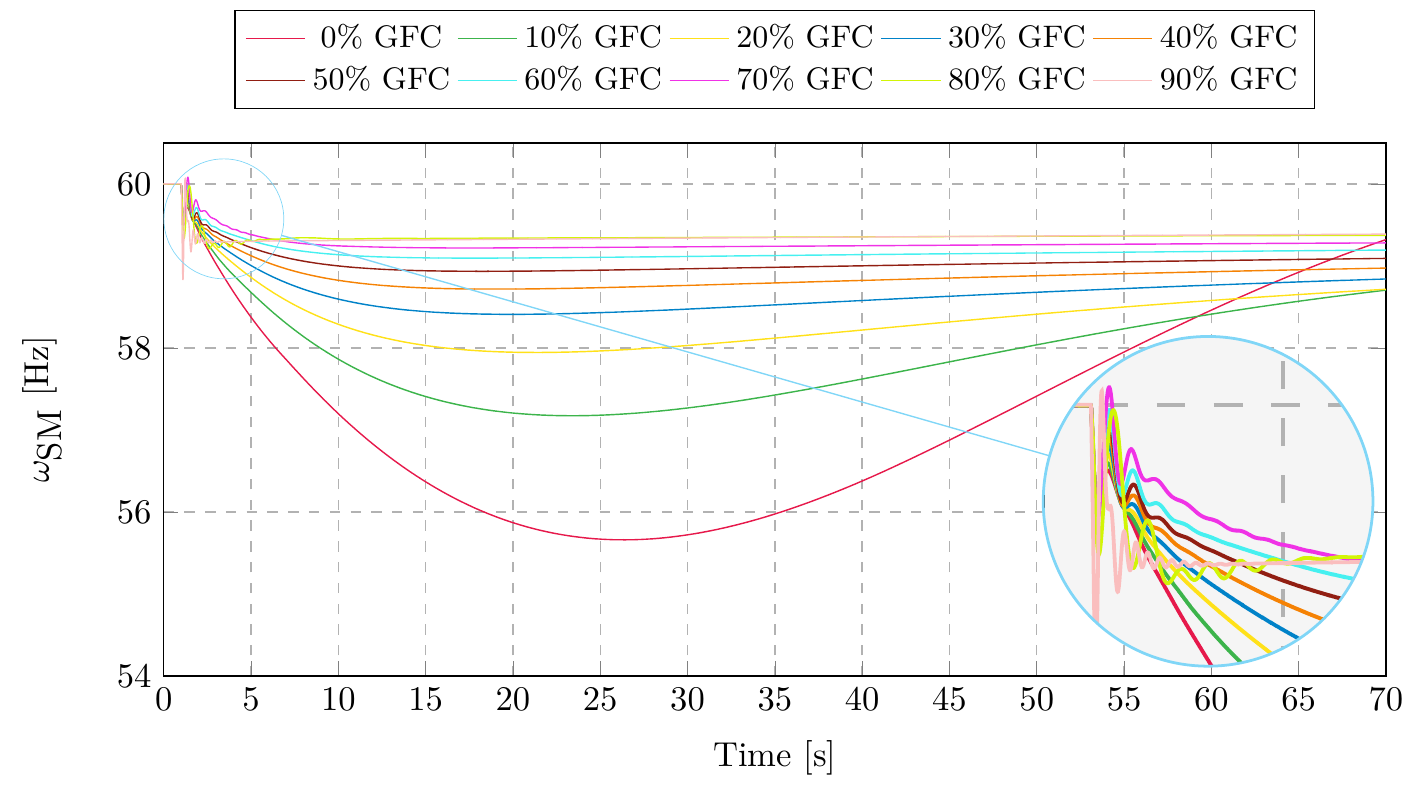}}
    \caption{Frequency evolution of the SM 2 following the loss of SM-GFC 1. When the GFC integration level is set to  80\% and 90\%, the PSS controllers for the remaining SMs have been identically retuned.}
    \label{fig:InteLevMag}
\end{figure}
\begin{figure}[t!]
    \centering
    {\includegraphics[width=0.42\textwidth]{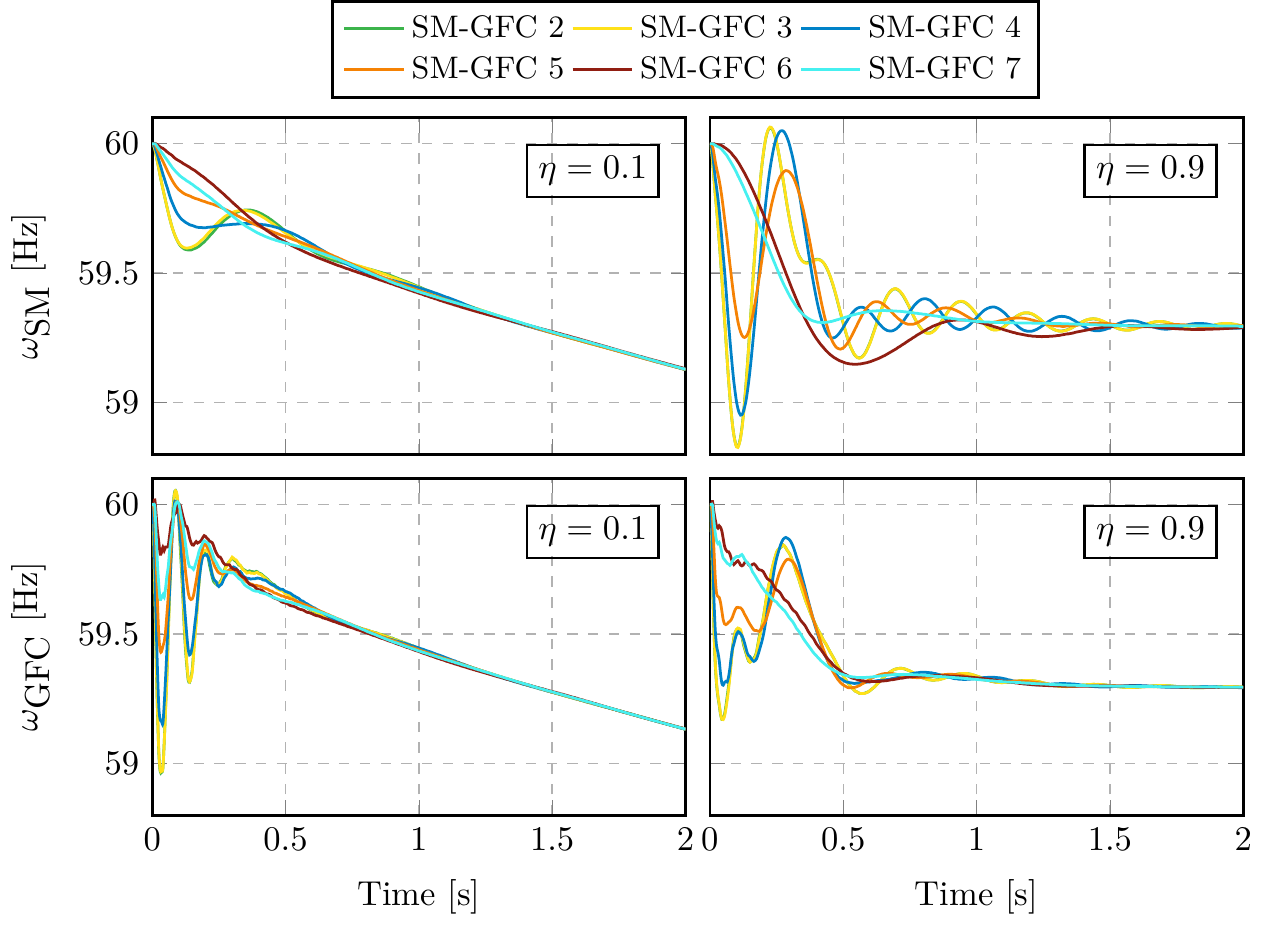}}
    \caption{The SM and GFC frequencies for two extreme integration levels, across the grid, following the loss of SM-GFC 1. Regardless of the integration level and the source of frequency signal (i.e., SM or GFC), units in the Northwest region - where contingency occurs - exhibit the largest oscillation.}
    \label{fig:FreqEvComp}
\end{figure} 
\subsection{Frequency performance under the worst contingency}\label{resultsFrequency}
Figure~\ref{fig:InteLevMag} illustrates the frequency time series of the SM 2 (the closest unit to the event) for the loss of the largest unit i.e., the SM-GFC 1 (see Figure~\ref{fig:Quebec}). The increasing integration of GFCs significantly improves the frequency nadir, but it degrades the RoCoF, when computed over a short time window (more on this topic in the next section). Moreover, time at which nadir occurs also is shortened. Although converters do not possess any significant inertia, their fast response curbs the impact of generator trip on the grid frequency. The behaviour for $80\%$ and $90\%$ is qualitatively different from the rest, due to the PSS retuning. The case of a pure converter-based grid is covered later in Section~\ref{resultsAllGFC}. Note that a similar behaviour has been observed under the other aforementioned grid-forming techniques.
\begin{remark}

\noindent By enforcing a slow frequency response for the GFCs - mimicking the slow turbine dynamics - GFCs can be made fully compatible with the time-scales of the SMs and their corresponding PSSs (i.e., reducing the time-scale separation of different generation units [Fig. 4] \cite{Uros2019}). However, fully mimicking the response of a SM would require to slow down the GFC frequency response artificially as well as significantly oversizing the GFCs. A much more viable solution is to adapt the PSS  parameters according to the penetration level.
\end{remark}
The time series in Figure \ref{fig:InteLevMag} correspond to the mechanical frequencies of the SM 2. For low values of $\eta$, these signals are expected to be representative of the bus frequencies across the grid. However, for a GFC-dominated grid, the GFC internal frequencies - being well-defined also in transients -  might be more descriptive of the frequencies across the grid. Figure~\ref{fig:FreqEvComp} illustrates the post-contingency frequency time series of SMs and GFCs, for the integration levels $\eta=0.1$ and $\eta=0.9$. As expected, the SM-GFCs 2-4 in the Northwest region (see Figure \ref{fig:Quebec}), which are closer to the event, exhibit the largest RoCoF values. Interestingly, at low penetration levels large oscillations appear at the GFCs before they synchronize with the SMs. At high integration levels we observe larger oscillations at the SMs. Further analyses are needed to conclude which set of signals is more relevant to describe the frequency behaviour for different integration levels. In any case - regardless of the integration level - the SMs mechanical frequencies are still needed to evaluate potential RoCoF-related issues associated with conventional generation.






\subsection{Evolution of the frequency metrics} \label{resultsnadirRocof}
\begin{figure}[t]
    \centering
    {\includegraphics[width=0.47\textwidth]{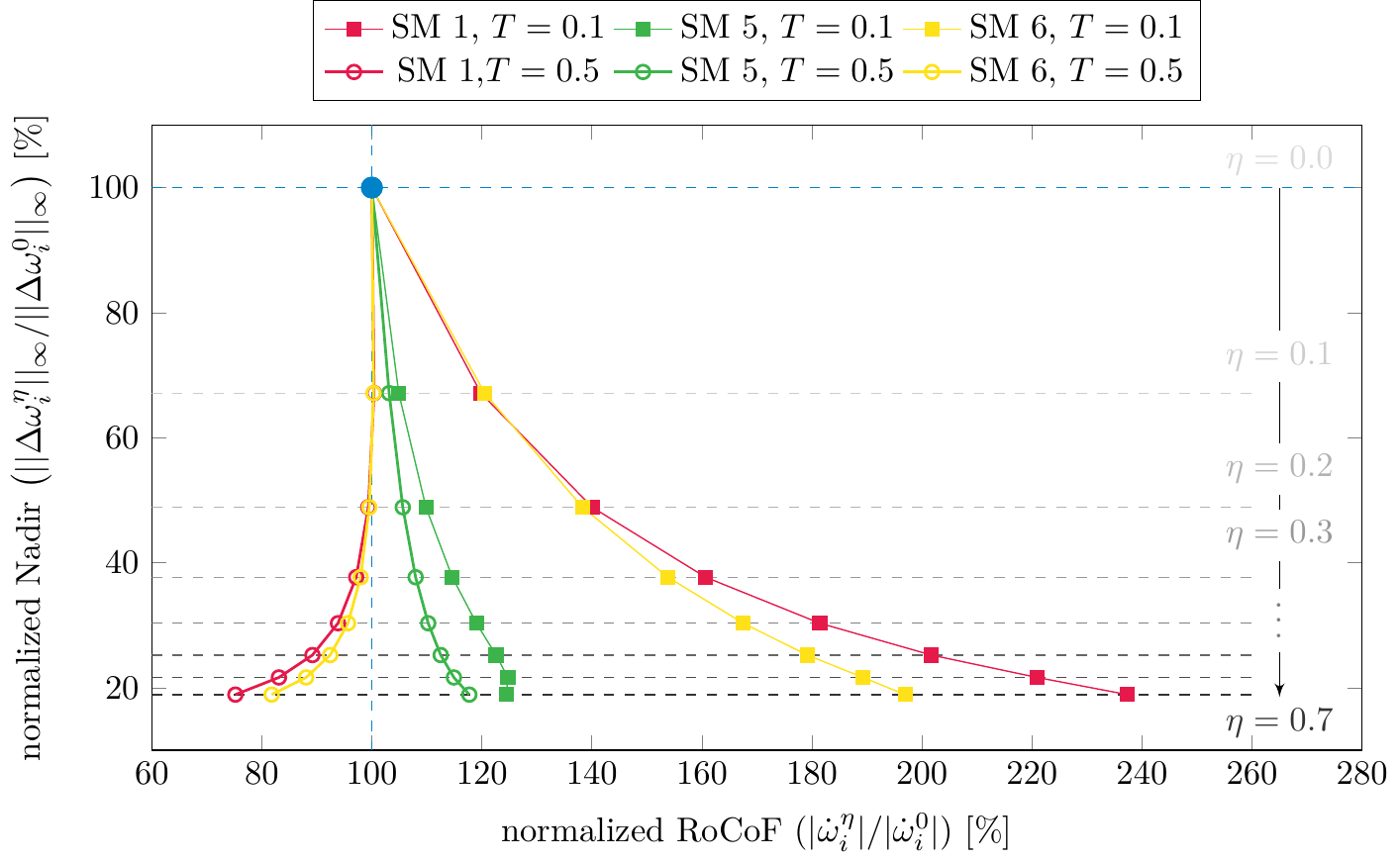}}
    \caption{Normalized frequency nadir vs RoCoF (calculated over two distinct time window) for different integration levels for all the SMs across the grid. The event corresponds to the loss of $2$ GW generation due to the disconnection of the HVDC link in Figure \ref{fig:Quebec}. All data points are normalized against the corresponding metric values of the baseline all-SMs system.}
    \label{fig:HVDCmetrics}
\end{figure}

While appropriate retuning of the PSS stabilizes the system, the results presented in the previous section suggest that the system dynamics drastically change depending on the GFCs integration level. To analyze and characterize this effect, we resort to the standard frequency stability metrics e.g., frequency nadir or maximum frequency deviation $||\Delta\omega||_\infty$ and  RoCoF(T) $|{\dot{\omega}}|$, formally defined for the generation unit $i$ as:
\begin{subequations}
\begin{align}
||\Delta\omega_i^\eta||_\infty &\coloneqq \max_{t\geq t_0} |\omega_i^\eta(t_0)-\omega_i^\eta(t)|,\\
|{\dot{\omega}_i^\eta}| &\coloneqq\frac{|\omega_i^\eta(t_0+T)-\omega_i^\eta(t_0)|}{T},
\end{align}
\end{subequations}
where $t_0$ is the time when the event occurs, $\omega_i^\eta$ the mechanical frequency at unit $i$ under penetration ratio $\eta$, and $T$ is the RoCoF calculation window. For ease of exposition, we consider in this subsection the HVDC link trip (see Figure \ref{fig:Quebec}), and evaluate these metrics based on the SMs frequencies across the grid. Figure \ref{fig:HVDCmetrics} depicts the frequency metrics evolution for SM $1$, $5$ and $6$ (representatives of each area in the grid) following the loss of $2$ GW generation caused by the HVDC link trip, for different values of $\eta$. We consider two RoCoF computation windows, namely $T_1=0.1 \mathrm{s}$ and $T_2=0.5 \mathrm{s}$ (i.e., computing RoCoF using different time windows), denoted as RoCoF($0.1$) and RoCoF($0.5$). Furthermore, the nadir and RoCoF values corresponding to a particular choice of $(\eta,T_{1,2})$ are normalized with respect to the metrics of the all-SMs system with the same RoCoF windows (i.e., $\eta=0$ and $T_{1,2}$). 
This removes the effect that RoCoF decreases when computed over a longer horizon. From Figure~\ref{fig:HVDCmetrics}, the following conclusions can be drawn:
\begin{itemize}
\item For the units SM 1 and 6 (the units far from the event), RoCoF($0.1$) deteriorates as $\eta$ increases, but  RoCoF($0.5$) improves with respect to the all-SM system. 

\item For the SM 5 - adjacent to the event - the RoCoF is less sensitive with respect to the integration level, since the collocated GFC reacts fast enough and comparable to the SM in the short term. 

\item In terms of absolute RoCoF values, i.e., not normalized against the all-SMs system's RoCoF, SM 5 is the one experiencing the largest RoCoF($0.1$) values, as expected (not shown here for space reasons).  

\item Similar observations were obtained in the previous subsection for the loss of SM-GFC 1, where the SMs in the same region (SM 2,3 and 4) exhibit the largest RoCoF($0.1$) values (see Figure~\ref{fig:FreqEvComp}). 
\end{itemize}
In other words, as inertia homogeneously decreases across the network, frequency decays faster right after a contingency, leading to larger RoCoF($0.1$) values. The GFCs respond slower than the instantaneous inertial response from SMs, but fast enough to arrest the frequency decay rate before $T=0.5s$, leading to smaller RoCoF($0.5$) values. 
\begin{remark} 
\noindent A similar analysis can be carried out using the GFC frequencies. There is no clear pattern on the evolution of the RoCoF($0.1$) for low values of $\eta$, since there are large oscillations within this time scale. In this case, the RoCoF($0.1$) metric is no longer insightful, and low values might hide large swings. It has been observed that RoCoF($0.5$) clearly decreases as $\eta$ increases.
\end{remark} 
\subsection{Discussion on the frequency metrics}\label{resultsDiscussion}
The presented results emphasize the relevance of the choice of the RoCoF window $T$, typically chosen to properly reflect frequency evolution, filter out noise and ignore fast transients, according to the characteristics of a grid \cite{entsoerocof_2017}. The presence of GFCs leads to new, fast dynamics and therefore the value of $T$ has to be reconsidered for the low-inertia systems. A natural reaction is to reduce the current choices for $T$ (typically between $500$ms and $1$s) to accommodate for the fast response of GFCs, but, as explained before, it can lead to misleading conclusions. On the other hand, large values of $T$ might be ineffective for protection devices, as dynamics are much faster under high values of $\eta$. High RoCoF values represent a challenge for existing settings of RoCoF relays, some load-shedding schemes, and conventional generation, that in general are not able to withstand sudden changes in speed and might disconnect to avoid damage.  Nonetheless, fast transients vanishing in less than $200$ms are not expected to be meaningful for the SM or RoCoF relays. Nonetheless, their influence on the grid-following converters can be significant, depending on the PLL implementations.

Notice as well how nadir is no longer uniform for all SMs under high penetration levels (e.g., see the time series corresponding to $\eta=0.9$ in Figure~\ref{fig:FreqEvComp}), caused by fast oscillations appearing adjacent to the event location and prior to the GFCs synchronization. For such a system, it might be needed to redefine the nadir metric to filter out these oscillations to obtain a meaningful metric which effectively reflects the severity of the grid contingency. Whether these fast dynamics need to be fully captured, ignored or just partially encapsulated in the metrics requires further in-depth investigations. This would depend on the effect of those fast dynamics across different components in the grid (grid-following devices, conventional generation, industrial loads, etc.). 
\subsection{All-GFC grid}
\label{resultsAllGFC}
\begin{figure}[t]
\centering
\includegraphics[width=.4\textwidth]{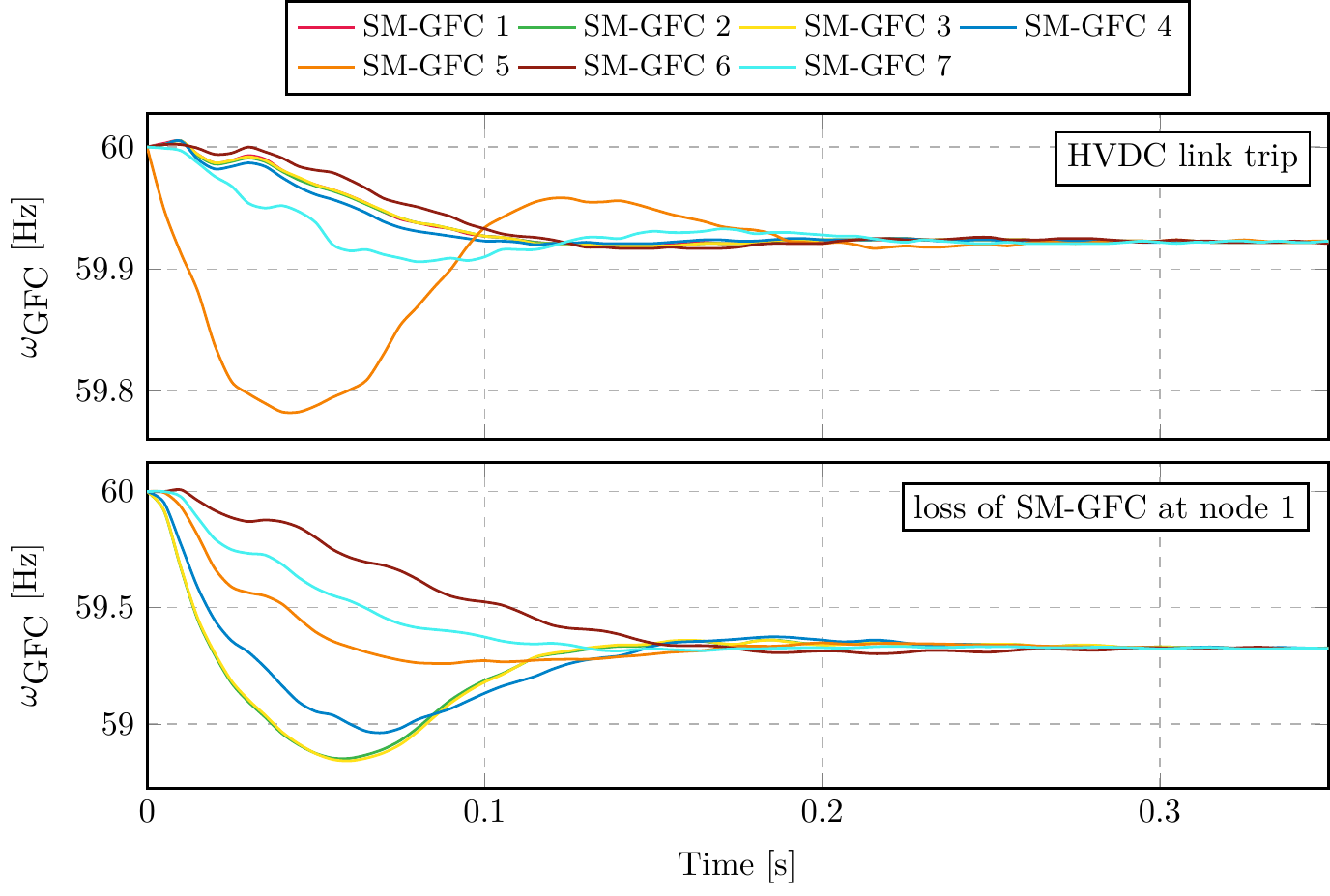}
\caption{Frequency time series for all units in an all-GFC grid.}
\label{allgfc_trip}
\end{figure}

We also explore a possible $100\%$ GFC scenario, without the presence of any SM. The controllers are tuned as in the previous section, that is, no modification has been carried out to stabilize the system. We compare in this case the trip of the HVDC link and the disconnection of generator 1. As shown in Figure~\ref{allgfc_trip}, after a very quick transient all converters synchronize under both contingencies, reaching a steady state before $300$ms. Once again, similar results have been observed under other grid-forming techniques, and combinations thereof.

Nadir is largely reduced in comparison to the all-SM grid, as the GFCs are orders of magnitude faster than the hydropower plants. For the case of the disconnection of generation unit 1, all generators in that area (2, 3 and 4) experience the largest values of short time-window RoCoF. On the other hand, for the other generators the response is nearly overdamped, and the nadir is equal to the steady state frequency deviation. This implies that nadir, as defined in~\eqref{metricsEq}, is much larger for those units close to the event. Unlike in SM-dominated grids, in all-GFC grids nadir can be reached before the generation units synchronize. Therefore, values are not uniform across all units in the grid, and depend largely on the location of the event. Similar conclusions can be reached for the disconnection of the HVDC link, as the RoCoF and nadir values for generator 5 are much larger than for the rest of the generators. 

These results question again the adequacy of the metrics in~\eqref{metricsEq} for converter-dominated grids. On one hand, large values of $T$ can render RoCoF useless as a metric, since the system might have reached a steady state\footnote{Assuming no secondary control or similar frequency-recovery scheme is implemented.}, and hence RoCoF would just be proportional to the droop coefficient of the grid. On the other hand, small values of $T$ that capture the first swing (around $50$ms for both events) are very impractical and sensitive to noise. Overall, it is unclear whether a metric is required to characterize these fast dynamics, whose effect in the grid might be questionable.
%
%
\section{Conclusions and outlook}
While GFCs have already been used at a microgrid scale, there exist serious doubts on the stability of large systems as GFCs replace SMs, especially at high penetration levels. This paper has explored the massive deployment of grid-forming converters and its effects on frequency behaviour. The presented results suggest that, under proper controller tuning, it is possible to guarantee frequency stability. Nonetheless, the grid dynamics change drastically, reaching steady state in the sub-second time range, orders of magnitude faster than the original pure-SM system. This has clear implications in terms of nadir and RoCoF, which might imply rethinking tuning of protection devices and load shedding schemes. There is also a need for PSS structures that can deal with a time-varying amount of inverter-based generation. To the best of our knowledge, no guidelines can be found for PSS tuning under high penetration scenarios.

Although in this work we have only covered the penetration of converters controlled as grid-forming units, it is expected that a large amount of devices will be operated as grid-following units. Large values of short time-window RoCoF might not be meaningful for frequency ride-through schemes in conventional generator or for RoCoF relays. However, grid following devices will try to synchronize to those fast transients, potentially creating large power transients. 

\bibliographystyle{IEEEtran}
\bibliography{IEEEabrv,pes2020_bib}

\end{document}